\documentclass[prl, floats, aps, prd, preprint, tightenlines, groupedaddress, nofootinbib, showpacs, byrevtex, superscriptaddress, preprintnumbers]{revtex4}

\usepackage{amsfonts}
\usepackage{amssymb,latexsym}
\usepackage{amsmath,amsbsy}
\usepackage{epsfig,bm}
\usepackage{graphicx,comment}
\def\cO{{\mathcal O}}
\def\cV{{\mathcal V}}

\begin{document}

\preprint{JLAB-THY-05-387}

\title{Vector Meson Mass Corrections at $ \cO(a^2) $ in PQChPT with\\
      Wilson and Ginsparg-Wilson quarks \\[25pt]}

\author{Hovhannes R. Grigoryan}
 \email{hgrigo1@jlab.org}
 \affiliation{Louisiana State University, Department of Physics \&
              Astronomy, 202 Nicholson Hall, Tower Dr., LA 70803, USA}
 \affiliation{Thomas Jefferson National Accelerator Facility, 12000
              Jefferson Ave., Newport News, VA 23606, USA}
 \affiliation{Laboratory of Theoretical Physics, JINR, Dubna, Russian
              Federation\\
 \qquad}

\author{Anthony W. Thomas}
 \email{awthomas@jlab.org}
 \affiliation{Thomas Jefferson National Accelerator Facility, 12000
              Jefferson Ave., Newport News, VA 23606, USA}




\begin{abstract}
We derive the mixed as well as unmixed lattice heavy meson chiral
Lagrangian up to order $ \cO(a^2) $, with Wilson and Ginsparg-Wilson
fermions. We consider two flavor partially quenched QCD and
calculate vector meson mass corrections up to order $ \cO(a^2) $,
including the corrections associated with the violation of
rotational $ O(4) $ symmetry down to the hypercubic group. Our
calculations also include the one-loop, phenomenological
contribution from the $\rho \rightarrow \pi \pi $ decay channel. The
final result is a chiral-continuum extrapolation formula with model
dependent coefficients from which one can recover the physical $
\rho $ meson mass from the large amount of current lattice data. As
a verification of our result, the chiral-continuum extrapolation
formula is compared with that used in numerical simulations.
\end{abstract}

\volumeyear{year}
\volumenumber{number}
\issuenumber{number}
\eid{identifier}
\startpage{1}
\endpage{8}

\maketitle

\section{Introduction}

In the present day lattice simulations the quark masses used are
much heavier than those in nature because of the increased
computational cost at low mass. Instead of trying to reach to
lighter quark masses one runs lattice simulations with heavier
quarks and then performs a chiral extrapolation to the physical
regime using the predictions of chiral perturbation theory (ChPT)
\cite{Weinberg, Gasser}. However, it is not totally consistent to
apply ChPT for extrapolation of lattice results because it is a
continuum theory which doesn't contain information about the lattice
spacing. That is why it is first required to extrapolate lattice
data to the continuum limit and then perform the chiral
extrapolation to physical regime. In order to consistently include
lattice discretization effects, Symanzik's effective theory needs to
be applied \cite{Symanzik:1983dc, Symanzik:1983gh}. This approach is
used, for example, in Refs.~\cite{Lee:1999zx, Aubin:2003mg,
Rupak:2002sm} and generalized in Refs.~\cite{Bar:2002nr, Bar:2003mh,
Tiburzi:2005vy} to study the mixed lattice theories with Wilson
fermions as sea and Ginsparg-Wilson fermions as valence quarks.
Being relatively expensive to simulate, Ginsparg-Wilson fermions
\cite{Ginsparg}, on the other hand, have an exact chiral symmetry
even on the lattice. The latter allows one to exploit the good
chiral properties of Ginsparg-Wilson fermions by considering them as
valence quarks\footnote{such valence fermions will have masses
smaller than the valence quark masses accessible using Wilson
fermions} and considering Wilson fermions as sea quarks
\cite{Neuberger:1997bg}. The Lagrangian approach used here is
necessary because of the large lattice spacing effects in unquenched
simulations (which means that the $ \cO(a^2) $ effects need to be
accounted). It represents an efficient method by
which to improve the numerical accuracy of the physical results.\\[-5pt]

In this paper we calculate the vector meson mass corrections up to
order $ \cO(a^2) $ in partially quenched chiral perturbation theory
(PQChPT) with graded symmetry $ SU(4|2) $, using Wilson and
Ginsparg-Wilson fermions and the following power counting scheme:
\begin{align}
p^2 \sim a \sim m_q
\end{align}
As the small parameters of our double expansion we will choose: $
\epsilon^2 = (a\Lambda, m_q/\Lambda) $, so that $ m_q \ll \Lambda
\ll 1/a $, (where $ \Lambda $ is a typical QCD scale).

To calculate the corrections coming from lattice discretization one
must formulate the corresponding lattice theory and match the new
operators with those in chiral perturbation theory (ChPT). The
standard procedure consists of the following parts:

\begin{enumerate}
\item[1.] writing the Lattice QCD action for the fermions one is going to use;
\item[2.] finding the corresponding Symanzik action up to a given order
in $ a $ based on symmetry constraints from underlying lattice
theory;
\item[3.] performing a spurion analysis with further projection to
ChPT.
\end{enumerate}

If we want to find the mass corrections to the vector mesons, we
need to formulate ChPT for both pseudoscalars and vector mesons.
ChPT for pseudoscalars with Wilson quarks was formulated, for
example, in Ref.\cite{Rupak:2002sm}. It was combined with
Ginsparg-Wilson fermions as a mixed theory in
Refs.\cite{Bar:2002nr,Tiburzi:2005vy}. Continuum heavy meson ChPT
was formulated in Ref.\cite{Jenkins:1995vb} and further developed
for QChPT and PQChPT in Refs.\cite{Booth:1996hk} and
\cite{Chow:1997dw}. An example of two-flavor PQChPT can be found in
Ref.\cite{Beane:2002vq}. The main disadvantage of similar heavy
vector meson ChPT formulations was that they didn't include terms
which explicitly break $ O(4) $ rotational symmetry down to the
hypercubic group. Recent studies, particularly by Tiburzi in
Ref.~\cite{Tiburzi:2005vy}, with Wilson and Ginsparg-Wilson quarks,
showed explicitly the terms in the heavy baryon Lagrangian that
break $ O(4) $ symmetry, up to order $ \cO(a^2) $. Following similar
lines, we formulate heavy meson partially quenched ChPT (HM PQChPT)
for Wilson and Ginsparg-Wilson fermions up to order $ \cO(a^2) $,
including chiral as well as continuum symmetry breaking.

The paper is organized in the following way: in Sec.II we review the
well developed PQChPT $ SU(4|2) $ Lagrangian for the pseudoscalar
meson sector with mixed and unmixed Wilson and Ginsparg-Wilson
fermions up to order $ \cO(a^2) $; in Sec.III the extended heavy
meson PQChPT Lagrangian is derived up to order $ \cO(a^2) $,
including the terms explicitly breaking $ O(4) $ rotational symmetry
down to the hypercubic group (this is actually the adjustment of
Ref.\cite{Tiburzi:2005vy} applied to vector mesons); in Sec.IV we
calculate the $ \rho $ meson mass corrections including
phenomenological one-loop correction from $ \rho \rightarrow \pi\pi
$ and in Sec.V the chiral-continuum extrapolation formula is derived
and compared with the phenomenological expression in
Ref.~\cite{Allton:2005fb}, in terms of which the $ \rho $ meson mass
was successfully extracted.

\section{Pseudoscalar Meson Sector}

Let us denote the quarks of $ SU(4|2) $ as
\begin{align}
Q = (x,y,u,d,\tilde{x},\tilde{y})^T,
\end{align}
where $ x $ and $ y $ label the valence fermions whose ghost
partners ($ \tilde{x},\tilde{y} $) cancel their effects in the sea
loops. Thus, the only sea loop contributions come from unquenched $
u $ and $ d $ quarks. Now, assume that $ m_x = m_y $ and $ m_u = m_d
$ which is natural if we want to calculate the correction to the
mass of the $ \rho $ meson. The mass and Wilson matrices \cite{SW}
are then:
\begin{align}
m_Q &= diag(m_x, m_x, m_u, m_u, m_x, m_x),\\
w_Q &= diag(w_v, w_v, w_s, w_s, w_v, w_v)
\end{align}
where $ w_v $ refers to valence and $ w_s $ to sea quarks. Note that
if quark is a Wilson fermion then: $ (w_Q)_i = 1 $, while if it is
GW then: $ (w_Q)_i = 0 $. This could be useful for the construction
of mixed types of ChPT, as in Refs.\cite{Bar:2002nr,
Tiburzi:2005vy}.

The leading order (LO) Lagrangian, $ L_2 \sim \cO(p^2, a, m_q) $,
which includes chiral symmetry breaking from quark mass and lattice
discretization, is given by the expression (\cite{Bar:2002nr,
Tiburzi:2005vy}):
\begin{align}
L_2 = \frac{f^2}{8}<\partial_{\mu}\Sigma\partial^{\mu}\Sigma^{\dag}>
- \ \lambda_m <m_Q\Sigma^{\dag} + m^{\dag}_Q\Sigma> - \ a\lambda_a
<w_Q\Sigma^{\dag} + w^{\dag}_Q\Sigma>
\end{align}
\noindent where the angular brackets stand for the super-trace over
the flavor indices, $ f $ is the pion decay constant, $
\lambda_{m(a)} $ are low energy constants and $ \Sigma = exp(2\imath
\Phi/f) $ contains the matrix of meson fields, $ \Phi \in SU(4|2) $.
The next to LO (NLO) Lagrangian would be of order $ \cO(\epsilon^4)
\sim \cO(p^4, p^2 a, p^2m_q, a^2, m^2_q, am_q) $, which is rather
lengthy (see Ref.\cite{Rupak:2002sm}). However, in calculating the
vector meson mass correction the NLO pseudoscalar meson Lagrangian
will give corrections of $ \cO(\epsilon^5) $ in which we are not
interested. We also explicitly omitted from the Lagrangian the
contribution from the $ \eta' $ (the $ SU(2) $ analog of it) which,
because of the $ U(1)_A $ anomaly, is integrated out from the theory
as a heavy degree of freedom. However, double hairpin vertices
involving the $ \eta' $ were used to find disconnected propagators.

The following results from Ref.\cite{Tiburzi:2005vy} are needed for
our calculations:
\begin{align}
m^2_{QQ'} &= \frac{4}{f^2}\left[\lambda_m (m_Q + m_{Q'}) +
a\lambda_a(w_Q + w_{Q'}) \right],\\[5pt]
G_{xy} &= - \frac{1}{2}\frac{(q^2 + m^2_{sea})}{(q^2 +
m^2_{val})^2}, \ \ \ \ \ \ \ G_{vs} = \frac{1}{q^2 + m^2_{mix}},
\end{align}
\noindent where $ m_{QQ'} $ is the mass of the pseudoscalar meson
with quarks $ Q $ and $ Q' $ and $ m_{sea} $, $ m_{val} $ and $
m_{mix} $ are mesons which consist of sea-sea, valence-valence and
valence-sea quarks correspondingly. In our case:
\begin{align}
m^2_{sea} &= \frac{8}{f^2}\left[\lambda_m m_u + a\lambda_a
\right],\\[5pt]
\nonumber m^2_{val} &=
\frac{8}{f^2}\left[\lambda_m m_x + a\lambda_a \right],\\[5pt]
\nonumber m^2_{mix} &= \frac{4}{f^2}\left[\lambda_m (m_x + m_u) + 2
a\lambda_a \right].
\end{align}
{}Finally, $ G_{xy} $ is the disconnected meson propagator, which
includes all diagrams with double hairpin insertions and $ G_{vs} $
is the connected propagator for the meson lines which
consist of one sea and one valence quark.\\

\section{Vector Meson Sector}

Standard chiral perturbation theory for vector mesons has been
formulated by Jenkins, Manohar and Wise in
Ref.~\cite{Jenkins:1995vb}, the quenched counterpart by Booth,
Chiladze and Falk in Ref.~\cite{Booth:1996hk} and the partially
quenched counterpart by C.~K.~Chow and S.~J.~Rey in
Ref.~\cite{Chow:1997dw}. The study of vector meson masses within the
context of ChPT\footnote{performing an expansion in terms of the
momenta, quark masses and $ 1/N_c $} can be found in
Ref.~\cite{Bijnens:1997ni}. Here we will construct the partially
quenched theory using mixed type of fermions and including the
lattice discretization effects up to order $ \cO(a^2) $.

The heavy vector meson Lagrangian, up to order $ \cO(\epsilon^4) $,
consists of the following parts:
\begin{align}
L^{(1)} &\sim \cO(p),\\
\nonumber L^{(2)} &\sim \cO(a, m_q), \\
\nonumber L^{(4)} &\sim \cO(a^2, m^2_q, am_q).
\end{align}
The Lagrangian $ L^{(3)} $ is omitted here because it contains
derivatives which can be eliminated using the equations of motion
\cite{Arzt:1993gz}. In the partially quenched (PQ) case the vector
meson multiplet (in the large $ N_c $ limit) is described by a $ 6
\times 6 $ matrix field (similar to the Goldstone meson sector):
\begin{align}
N_{\mu} = \left(
            \begin{array}{ccc}
              \cV & \Psi \\
      \Psi^\dag & \tilde{\cV}\\
            \end{array}
          \right)_{\mu}
\end{align}
\noindent where $ \cV $ is a $ (4 \times 4) $ matrix of $ q\bar{q} $
states, $ \tilde{\cV} $ is $ (2 \times 2) $ matrix which consists of
$ \tilde{q}\bar{\tilde{q}} $ states, $ \Psi $ and $ \Psi^\dag $ are
rectangular matrices involving fermionic bound states with $
q\bar{\tilde{q}} $ or $ \tilde{q}\bar{q} $ content. Under the $ G =
SU(4|2)_L \times SU(4|2)_R \times U(1)_V $ graded chiral symmetry
group, the heavy meson field transforms as:
\begin{align}
N_{\mu} \rightarrow U N_{\mu} U^\dag, \ \  U \in G
\end{align}
\noindent and under charge conjugation:
\begin{align}
CN_{\mu}C^{-1} = - N^\textit{T}_{\mu}.
\end{align}
As in conventional ChPT \cite{Jenkins:1995vb} (as well as QChPT
\cite{Booth:1996hk} and PQChPT \cite{Chow:1997dw}), the vector meson
multiplet will be treated as heavy, static source. We will also
assume that the static vector meson propagates with a fixed
four-velocity $ v^{\mu} $, $ v^2 = 1 $ and interacts with soft
Goldstone multiplets. The three polarization states of the vector
mesons are perpendicular to the propagation direction (i.e. $ v
\cdot N = 0 $).

Taking all of these into account, let us now proceed to construct
the relevant Lagrangians. The LO heavy-meson chiral Lagrangian will
consist of terms of order $ \cO(p) \sim \cO(\epsilon) $ which are:
\begin{align}\label{kin}
L_{kin} = - \imath <N^\dag_{\mu}(v\cdot D) N_{\mu}> - \  \imath A_1
<N^\dag_{\mu}>(v\cdot D) < N_{\mu}> \  \  \sim \cO(\epsilon),
\end{align}
\begin{align} \label{int}
L_{int} = \imath
g_1<N^\dag_{\mu}><N_{\nu}A_{\lambda}>v_{\sigma}\epsilon^{\mu\nu\lambda\sigma}
+ h.c. + \imath g_2
<\{N^\dag_{\mu},N_{\nu}\}A_{\lambda}>v_{\sigma}\epsilon^{\mu\nu\lambda\sigma}
+
\\[5pt] \nonumber
\imath g_3
<N^\dag_{\mu}><N_{\nu}><A_{\lambda}>v_{\sigma}\epsilon^{\mu\nu\lambda\sigma}
+ \imath g_4
<N^\dag_{\mu}N_{\nu}><A_{\lambda}>v_{\sigma}\epsilon^{\mu\nu\lambda\sigma}
\sim \cO(\epsilon),
\end{align}
\noindent where the following notations are implied:
\begin{align}
D_{\mu} =& \partial_{\mu} + [V^{\mu},\ \ ], \ \ \ V^{\mu} =
\frac{1}{2}\left(\xi \partial^{\mu}\xi^\dag + \xi^\dag
\partial^{\mu}\xi \right), \ \ \ A^{\mu} =
\frac{1}{2}\left(\xi \partial^{\mu}\xi^\dag - \xi^\dag
\partial^{\mu}\xi \right) \\[5pt]
 \ \ \ \ \ \ &\  \ \ \ \ \ \ \ \ \ \ \ \ \ \ \xi = \sqrt{\Sigma}, \ \ \  M = \frac{1}{2}\left(\xi^{\dag} m_Q
\xi^{\dag} + \xi m_Q \xi \right).
\end{align}
Here we treat $ A_1 $ as a small quantity, so the effects of this
term can be reabsorbed in the other parameters by the field
redefinition:
\begin{align}
 N'_{\mu} = N_{\mu} + \frac{A_1}{2} <N_{\mu}>.
\end{align}
This redefined field still transforms under the chiral group in the
same way as $ N_{\mu} $. As we may notice, there are four-types of
chiral invariant interactions in Eq.~(\ref{int}) (consistent with
graded chiral symmetry) between vector meson multiplets and
pseudoscalar mesons. In the large $ N_c $ limit the coupling
constants scale with $ N_c $ as follows: $ g_1 \sim \frac{1}{N_c} $,
$ g_2 \sim 1 $, $ g_3 \sim \frac{1}{N^2_c} $, $ g_4 \sim
\frac{1}{N_c} $. However, according to Ref.~\cite{Booth:1996hk}, the
corresponding quark (and color) flow diagrams are not planar, so
they need to be multiplied by an additional factor of $ 1/N_c $.

Because we are interested in finding the one-loop correction to the
flavor charged vector meson mass in isospin limit there will be no
contribution from the terms proportional to $ g_1 $ and $ g_3 $.
These terms are responsible for quark flow diagrams with
disconnected vector meson line, and because we require to have two
different flavors at valence lines we cannot have this type of
hairpins (as we don't consider any flavor changing interactions).
The diagrams from the term proportional to $ g_{4} $ are one hairpin
like diagrams (due to coupling of the vector meson with the two
flavor analog of $ \eta' $). Sum of these diagrams with vacuum
bubble insertions can be shown to vanish after the decoupling of
flavor singlet\footnote{For example, in Ref.~\cite{Chow:1997dw} it
is assumed that $ g_{1,3,4} = 0 $ and $ g_2 = 0.75 $. In principle,
our considerations are more general and we could leave these
couplings as free parameters which are to
be determined by further lattice simulations.}.\\

The next to LO (NLO) Lagrangian -- $ L_2 \sim \cO(\epsilon^2) $,
according to our counting scheme, will consist of terms of order $
\cO(m_q) $ and $ \cO(a) $.  The first Lagrangian $ L_{mass} \sim
\cO(m_q) $ is:
\begin{align} \label{mass}
\nonumber L_{mass} &= \bar{\mu}<N^\dag_{\mu}N_{\mu}> + \  \mu_1
<N^\dag_{\mu}><N_{\mu} > + \\[5pt] \nonumber &+ \ a_1<N^\dag_{\mu}N_{\mu}>< M
> + \ a_2 <N^\dag_{\mu}><N_{\mu} >< M > \\[5pt] &+
\lambda_1\left(<N^\dag_{\mu}><N_{\mu}M > + h.c.\right) +
\lambda_2<\{N^\dag_{\mu},N_{\mu}\}M > \sim  \cO(\epsilon^2),
\end{align}
\noindent where the first term in Eq.~(\ref{mass}) corresponds to
'residual mass' $ \bar{\mu} $ of vector meson multiplets which is
always possible to remove by a suitable reparametrization
transformation \cite{Falk:1992fm}, (through the choice of the
reference momentum). Here we will use this freedom and assign $
\bar{\mu} = 0 $. The second term is small because of Zweig's rule,
therefore we treat it as the same order as the quark masses, $
\cO(\epsilon^2) $ \cite{Bijnens:1997ni}. The chirally invariant
terms in Eq.~(\ref{mass}) proportional to $ a_{1,2} $ are
contributing directly to the vector meson mass. For further
simplicity and in order to match the conventional ChPT in
Ref.~\cite{Jenkins:1995vb} in continuum and large $ N_c $ limits we
will assign $ a_{1,2} = 0 $. The last two terms in $ L_{mass} $,
proportional to $ \lambda_{1,2} $, correspond to $ SU(2) $ isospin
breaking because of the quark masses. To match with the mass
Lagrangian in Ref.~\cite{Jenkins:1995vb} (Eq.24) we will take $
\mu_1 = \lambda_1 = 0 $ throughout this paper reminding that this
particular choice will be justified only by the final results.

From the analysis above it follows that the flavor non-diagonal heavy
vector meson propagator is:\\[-25pt]

\begin{align}
G_{\mu \nu} = \frac{(v^{\mu}v^{\nu} - g^{\mu\nu} )}{v \cdot k +
\imath \epsilon}
\end{align}

\noindent where $ k^{\mu} = p^{\mu}_V - ( M_V - \bar{\mu} )v^{\mu} $
is its residual momentum, $ p^{\mu}_V $ and $ M_V $ are the four-momentum
and mass of the vector meson correspondingly.\\

The second NLO term $ L_a \sim \cO(a) $ could be written in the
following form:
\begin{align}
L_{a} &= \alpha_1 \left(<N^\dag_{\mu}><N_{\mu}W_{+}> + \ h.c.\right)
+ \ \alpha_2<\{N^\dag_{\mu},N_{\mu}\}W_{+}> \\ \nonumber &+ \
\alpha_3 <N^\dag_{\mu}><N_{\mu}><W_{+}> + \
\alpha_4<N^\dag_{\mu}N_{\mu}><W_{+}> \ \ \sim \cO(\epsilon^2),
\end{align}
\noindent with the definition
\begin{align}
W_{+} \equiv \frac{a\Lambda^2}{2}\left(\xi^{\dag} w_Q \xi^{\dag} +
\xi w_Q \xi\right),
\end{align}
\noindent where $ \alpha_{1,2,3,4} $ are undetermined low energy
constants. Notice that the terms in $ L_a $ break chiral symmetry in
the same way as masses and that the dominant (with respect to $ N_c
$) term is one proportional to $ \alpha_2 $. The terms proportional
to $ \alpha_{1,3} $ don't give any contribution to the vector meson
mass because (similar to $ g_{1,3} $) these interactions give
hairpin insertions in vector meson line.

Now, let us write the next to NLO (NNLO) Lagrangian of order $
\cO(\epsilon^4) $. As we mention in the beginning the NNLO
Lagrangian will consists of the following parts: $ L_{am_q} \sim
\cO(am_q) $, $ L_{m^2_q} \sim \cO(m^2_q) $ and $ L_{a^2} \sim
\cO(a^2) $.\\[-5pt]

Using the properties of $ M $ and $ W_{+} $ under chiral
transformations the first term is:
\begin{align}
L_{am_q} &= \beta_1 <N^\dag_{\mu}><N_{\mu}><W_{+}><M> + \ \beta_2
<N^\dag_{\mu} N_{\mu}><W_{+}><M> \\ \nonumber &+ \ \beta_3
<N^\dag_{\mu}><N_{\mu}><W_{+}M> + \ \beta_4\left( <N^\dag_{\mu}
W_{+}><N_{\mu}><M> + h.c \right) \\ \nonumber &+ \ \beta_5
\left(<N^\dag_{\mu} M><N_{\mu}><W_{+}> + h.c \right) + \ \beta_6
<N^\dag_{\mu} N_{\mu}><W_{+}M> \\ \nonumber &+ \ \beta_7
\left(<N^\dag_{\mu} W_{+}><N_{\mu}M> + h.c \right) + \ \beta_8
<\{N^\dag_{\mu}, N_{\mu}\} W_{+}><M> \\ \nonumber &+ \ \beta_9
<\{N^\dag_{\mu}, N_{\mu}\} M ><W_{+}> + \ \beta_{10}
\left(<N^\dag_{\mu}><N_{\mu} \{W_{+}, M\}> + h.c \right) \\
\nonumber &+ \ \beta_{11} <\{N^\dag_{\mu}, \{N_{\mu}, W_{+}\}\}M >
\end{align}
Similarly, the second term has the structure:
\begin{align}\label{WW}
L_{m^2_q} &= r_1<N^\dag_{\mu}><N_{\mu}>< M >< M > + \
r_2<N^\dag_{\mu}><N_{\mu}>< MM > \\ \nonumber & + \ r_3<N^\dag_{\mu}
N_{\mu}>< M >< M > + \ r_4 \left(<N^\dag_{\mu} W_{+}><N_{\mu}>< M >
+ h.c. \right) \\ \nonumber & + \ r_5<N^\dag_{\mu} N_{\mu}>< MM > +
\ r_6<N^\dag_{\mu} M ><N_{\mu} M > \\ \nonumber &+ \
r_7<\{N^\dag_{\mu}, N_{\mu}\} M >< M > +
r_8\left(<N^\dag_{\mu}><\{N_{\mu},M\} M > + h.c. \right) \\
\nonumber &+ \ r_{9}<\{N^\dag_{\mu}, \{N_{\mu},M\}\}M >,
\end{align}
\noindent where $ \beta_{1,...11} $ and $ r_{1,...,9} \sim
1/\Lambda_1 $ are undetermined constants ($ \Lambda_1 $ is some
energy scale). The terms \ $ r_{1,...,8} $ \ are $ 1/N_c $
suppressed, according to Ref.~\cite{Bijnens:1997ni}. These
Lagrangians give only tree level contributions to the mass at $
\cO(\epsilon^4) $ which are proportional to $ am_q $ and $ m^2_q $
correspondingly.

The NNLO Lagrangian of order $ \cO(a^2) $ will consist of four
parts, the first part has the form:
\begin{align}\label{WW}
L^{WW}_{a^2} &= q_1<N^\dag_{\mu}><N_{\mu}><W_{+}><W_{+}> + \
q_2<N^\dag_{\mu}><N_{\mu}><W_{+}W_{+}> \\ \nonumber & + \
q_3<N^\dag_{\mu} N_{\mu}><W_{+}><W_{+}> + \ q_4 \left(<N^\dag_{\mu}
W_{+}><N_{\mu}><W_{+}> + h.c. \right) \\ \nonumber & + \
q_5<N^\dag_{\mu} N_{\mu}><W_{+} W_{+}> + \ q_6<N^\dag_{\mu}
W_{+}><N_{\mu} W_{+}> \\ \nonumber &+ \ q_7<\{N^\dag_{\mu},
N_{\mu}\} W_{+}><W_{+}> + q_8\left(<N^\dag_{\mu}><\{N_{\mu},W_{+}\}
W_{+}> + h.c. \right) \\ \nonumber &+ \ q_{9}<\{N^\dag_{\mu},
\{N_{\mu},W_{+}\}\}W_{+}>,
\end{align}
\noindent where $ q_{1,...,9} \sim - 1/\Lambda $.

In order to construct the second and third parts of $ \cO(a^2) $
Lagrangian, taking into account the relation $ \bar{w}_Q \equiv 1 -
w_Q $ (from \cite{Tiburzi:2005vy}), let us define:
\begin{align}
 P_{+} \equiv \frac{1}{2}\left[\xi^{\dag} (w_Q - \bar{w}_Q) \xi + \xi (w_Q - \bar{w}_Q) \xi^{\dag} \right]
\end{align}
\noindent which transforms under $ G $ as $ UP_+U^{\dag} $. Using
this spurion field the second part of the $ \cO(a^2) $  Lagrangian
will be:
\begin{align}
L^P_{a^2} &= \gamma_0 <N^\dag_{\mu} N_{\mu}> + \gamma_1
<N^\dag_{\mu}><N_{\mu}><P_{+}> + \gamma_2 <N^\dag_{\mu}
N_{\mu}><P_{+}> \\[5pt] \nonumber &+ \gamma_3 <\{N^\dag_{\mu},
N_{\mu}\}P_{+}> + \gamma_4\left(<N^\dag_{\mu} ><N_{\mu} P_+> +
h.c.\right),
\end{align}
\noindent where $ \gamma_{0,...,4} \sim - a^2\Lambda^3/\Lambda_1 $.
Similar to Eq.~(\ref{WW}), the third part of $ \cO(a^2) $ Lagrangian
will be written in the form:
\begin{align}
L^{PP}_{a^2} &= \delta_1<N^\dag_{\mu}><N_{\mu}><P_{+}><P_{+}> + \
\delta_2<N^\dag_{\mu}><N_{\mu}><P_{+}P_{+}> \\ \nonumber & + \
\delta_3<N^\dag_{\mu} N_{\mu}><P_{+}><P_{+}>
 + \ \delta_4 \left(<N^\dag_{\mu} P_{+}><N_{\mu}><P_{+}> + h.c.
 \right) \\ \nonumber &
+ \ \delta_5<N^\dag_{\mu} N_{\mu}><P_{+} P_{+}> + \
\delta_6<N^\dag_{\mu} P_{+}><N_{\mu} P_{+}> \\ \nonumber &+
\delta_7<\{N^\dag_{\mu}, N_{\mu}\} P_{+}><P_{+}> + \
\delta_8\left(<N^\dag_{\mu}><\{N_{\mu},P_{+}\} P_{+}> + h.c. \right)
\\ \nonumber & + \ \delta_{9}<\{N^\dag_{\mu},
\{N_{\mu},P_{+}\}\}P_{+}>,
\end{align}
\noindent where $ \delta_{1,...,9} \sim - a^2\Lambda^3/\Lambda_1 $.

{}Finally, following Ref.~\cite{Tiburzi:2005vy}, we write the last
part of $ \cO(a^2) $ Lagrangian, which explicitly break $ O(4) $
symmetry, in the form:
\begin{align}
\nonumber L^{c}_{a^2} &= k_1 v^4_{\mu} <N^\dag_{\nu} N_{\nu}> + \
k_2 v^4_{\mu} <N^\dag_{\nu}\{N_{\nu}, P_{+}\}> + \ k_3v^4_{\mu}
<N^\dag_{\nu} N_{\nu}><P_{+}> \\[5pt] &+ \ k_4 v^2_{\mu}<N^\dag_{\mu}
N_{\mu}> + \ k_5 v^2_{\mu}<N^\dag_{\mu}\{N_{\mu}, P_{+}\}> + \ k_6
v^2_{\mu}<N^\dag_{\mu} N_{\mu}><P_{+}>,
\end{align}
\noindent where $ k_{1,...,6} \sim -a^2 \Lambda^3/\Lambda_1 $. Later
we will see that for the vector meson at rest the mass corrections
from lattice discretization up to $ \cO(\epsilon^4) $ are of the
form: $ M_{\rho}(a) = M_{\rho}(0) - a^2\Lambda D $, where $ D $ is
some constant.

\section{Vector Meson Masses up to order $ \cO(a^2) $}

This developed heavy meson (HM) PQChPT with $ SU(4|2) $ graded
symmetry allows us to calculate the corresponding vector meson mass
correction. The vector meson mass in both chiral and continuum
expansions can be written in the form:
\begin{align}
M^2_V = \left( M_0(\mu) + \sigma^{(2)}_{tree}(\mu) +
\sigma^{(3)}_{tree}(\mu) + \sigma^{(4)}_{tree}(\mu) + ... \right)^2
+ \sigma^{(3)}_{loop}(\mu) + \sigma^{(4)}_{loop}(\mu) \ ...,
\end{align}
\noindent where $ M_0(\mu) $ is the renormalized vector meson mass
in continuum and chiral limits, $ \sigma^{(n)}_{tree(loop)}(\mu) $
is the tree (loop) contribution to the self energy of order $
\cO(\epsilon^n) $ and $ \mu $ is the renormalization scale. At order
$ \cO(\epsilon^2) $ we have the following tree level contribution to
the mass:
\begin{align}
\sigma^{(2)}_{tree} = - 2\lambda_2 m_x - 2a\Lambda^2
\left[\alpha_2w_v + \alpha_4w_s\right].
\end{align}
We will not have any contribution of order $ \cO(\epsilon^3) $
simply because we don't have Lagrangian of this order, so $
\sigma^{(3)}_{tree}(\mu) = 0 $.

Now, let us write the $ \cO(\epsilon^4) $ correction as:
\begin{align}
\nonumber \sigma^{(4)}_{tree} &= - \frac{1}{\Lambda_1}\left[
a\Lambda^2
\left(A m_x + Bm_u\right) + \bar{A} m^2_x + \bar{B} m^2_u \right] \\
 & \  \  \  \ \ - \ a^2\Lambda^3\left(C + Dv^4_{\mu}N^{\dag}_{\nu}N_{\nu}
+ \bar{D}v^2_{\mu}N^{\dag}_{\mu}N_{\mu} \right),
\end{align}
\noindent where $ A $, $ B $, $ \bar{A} $, $ \bar{B} $, $ C $, $ D $
and $ \bar{D} $ are undetermined coefficients \footnote{ for
example, $ A \equiv \left(2\beta_5 + 2\beta_9 \right)w_s +
\left(2\beta_7 + 4\beta_{10} + 4\beta_{11}\right)w_v $ and $ B
\equiv \left(\beta_1 + \beta_2 + \beta_3 + \beta_6 \right)w_s +
\left(2\beta_4 + 2\beta_8 \right)w_v $}.

The contribution from the tadpole-like loops, with double hairpin
insertion is:
\begin{align}
\sigma^{(4)}_{loop} = \frac{4\alpha_2a\Lambda^2}{\Lambda^2_1}&\left[
(w_s + w_v)F(m_{mix}, \mu) - 2w_v F(m_{val}, \mu) - 2w_v\Delta
m^2\frac{\partial
F(m_{val}, \mu)}{\partial m^2_{val}}\right]\\[5pt]
\ \ \ \ \ \ \  &\ \ \ \ \ \ \ \ \ \ \ \ \ \ \ \ \ \ F(m, \mu) \equiv
m^2 ln \frac{m^2}{\mu^2}
\end{align}
\noindent where $ \Delta m^2 \equiv m^2_{val} - m^2_{sea} =
\frac{8}{f^2}(m_x - m_u) $. The corrections $ \sigma^{(3)}_{loop} $
were calculated in Ref.~\cite{Allton:2005fb}, including the  loop
contribution from $ \rho \rightarrow \pi\pi $ (which is included
phenomenologically). Rather than repeating the expressions used in
that analysis of data over a wide range of meson mass, which were
based on finite range regularization, we show just the leading or
next-to-leading non-analytic behavior.

\section{Chiral Expansion Formula}

The complete chiral expansion for the vector meson mass at rest is
then:
\begin{align}
M^2_V =& \left( M_0(\mu) - \chi_1 m_x -  a\Lambda^2\chi_2 m_u -
\bar{\chi}_1m^2_x -  \bar{\chi}_2m^2_u -
a\Lambda^2 \left[\alpha_2w_v + \alpha_4w_s\right] + c_b a^2 \right)^2 \\[8pt] \nonumber &+ \
 \frac{4\alpha_2a\Lambda^2}{\Lambda^2_1}\left[ (w_s +
w_v)F(m_{mix}, \mu) - 2w_v F(m_{val}, \mu) - 2w_v \Delta m^2
\frac{\partial F(m_{val}, \mu)}{\partial m^2_{val}}\right] +
\Sigma^{tot}(\mu),
\end{align}
\noindent where $ \chi_{1} = 2\lambda_2 + Aa\Lambda^2/\Lambda_1 $, $
\chi_{2} = B/\Lambda_1 $, $ \bar{\chi}_1 = \bar{A}/\Lambda_1 $, $
\bar{\chi}_2 = \bar{B}/\Lambda_1 $ and $ c_b = - \Lambda^3 D' $ are
new parameters as functions of the previous ones. The term in square
brackets comes from tadpole diagrams and $ \Sigma^{tot} $ is the
total one-loop correction from Ref.\cite{Allton:2005fb} which
includes the $ \sigma^{(3)}_{loop}(\mu) $ contribution. If one fixes
the masses of the sea quarks and introduces new dimensionful
(reparametrized) extrapolation coefficients, then we will have:
\begin{align}
M^2_V =& \left( c_0 - \Lambda^2(\alpha_2 w_v + \alpha_4 w_s + \chi_2
m_u)a + c_b a^2
+ (\chi + c a) m^2_{\pi} + \bar{\chi}m^4_{\pi} \right)^2 \\[5pt]
\nonumber - & \ \frac{4\alpha_2a\Lambda^2}{\Lambda^2_1}\left\{(w_s +
w_v)F(m_{mix}) - 2w_v F(m_{val}) - 2w_v \Delta m^2 \frac{\partial
F(m_{val})}{\partial m^2_{val}} \right\} + \Sigma^{tot}(\mu),
\end{align}
\noindent where $ m^2_{\pi} = \frac{8\lambda_m}{f^2} m_x $ and $ c_0
$, $ c $, $ \chi $, and $ \bar{\chi} $ are new independent
parameters which are functions of the old ones.

Now, consider two different cases following from the theory:
\begin{enumerate}
\item[1.] when the sea and valence quarks are Wilson-type fermions then the expansion
          takes the form:
\begin{align}
M^2_V =& \left( c_0 - \Lambda^2(\alpha_2 + \alpha_4 + \chi_2 m_u)a +
c_b a^2
+ (\chi + c a) m^2_{\pi} + \bar{\chi}m^4_{\pi} \right)^2 \\[5pt]
\nonumber - & \
\frac{8\alpha_2a\Lambda^2}{\Lambda^2_1}\left\{F(m_{mix}) -
F(m_{val}) - \Delta m^2 \frac{\partial F(m_{val})}{\partial
m^2_{val}} \right\} + \Sigma^{tot}(\mu),
\end{align}
\item[2.] when the valence quarks are Ginsparg-Wilson fermions and the sea quarks
          are Wilson fermions we have:
\begin{align}
M^2_V =& \left( c_0 - \Lambda^2(\alpha_4 + \chi_2 m_u)a + c_b a^2
+ (\chi + c a) m^2_{\pi} + \bar{\chi}m^4_{\pi} \right)^2 \\[5pt]
\nonumber - & \ \frac{4\alpha_2a\Lambda^2}{\Lambda^2_1}F(m_{mix})+
\Sigma^{tot}(\mu),
\end{align}
\end{enumerate}
The analysis in Ref.\cite{Allton:2005fb}, which corresponds to the
first case, was not sensitive to the existence of tadpole-like
diagrams which are proportional to $ \alpha_2 $, so there is a good
reason to believe that $ \alpha_2 $ is consistent with zero.
Similarly, $ c $ and $ \alpha_4 + \chi_2 m_u $ or $ \alpha_4 $ and $
\chi_2 $ are also consistent with zero\footnote{$ \alpha_4 + \chi_2
m_u = 0 $ can be achieved if $ \alpha_4 = \chi_2 = 0 $ because $ m_u
$ is a variable.} according to the analysis in the same reference.
Comparison with the numerical simulations in
Ref.~\cite{Allton:2005fb} tells us that $ \bar{\chi} \approx - 0.061
GeV^{-3} $ which gives a negligible contribution to the vector meson
mass close to chiral limit.

\section{Conclusion}

Based on the symmetries of the mixed (as well as unmixed) lattice
QCD Lagrangian with Wilson and Ginsparg-Wilson fermions, heavy
vector meson, partially quenched chiral perturbation theory was
derived with explicit breaking of continuum symmetry up to order $
\cO(a^2) $. Using this formalism for the two flavor case, we
extracted chiral and continuum expansion formulas with model
dependent inputs with further comparison to the recently published
numerical simulation which incorporated data over a wide range of
meson masses.

We perform our calculations in the isospin symmetry limit when $ m_x
= m_y $ and $ m_u = m_d $. In our calculations of vector meson
masses we include the phenomenological contribution from the $\rho
\rightarrow \pi \pi $ decay which can't be consistently extracted
from heavy meson ChPT (because this is strongly relativistic
process). We also didn't include the finite volume effects from the
lattice.

The comparison with the extrapolation expression successfully used
in Ref.~\cite{Allton:2005fb} leads us to the conclusion that in our
approximation chiral and continuum symmetries are mostly broken by
the operators in $ L_{mass} $ and $ L_{a^2} $ and it is consistent
to disregard the contributions from the Lagrangians: $ L_{a} $, $
L_{m^2_q} $ and $ L_{am_q} $. This also gives us a simplified
chiral-continuum extrapolation formula for further simulations using
Wilson sea and Ginsparg-Wilson valence quarks.

This study is important because it allows us to check the validity
of lattice QCD simulations, QCD and ChPT as well, by performing
systematic chiral and continuum expansions of lattice data to the
real world with further comparison to the experimental data.

\section*{Acknowledgements}

H.R.G. would like to thank Jerry P. Draayer for his support, the
Southeastern Universities Research Association (SURA) for a Graduate
Fellowship and Louisiana State University for a Scholarship
partially supporting his research. He would also like to acknowledge
S. Vinitsky and G. Pogosyan for support at the Joint Institute of
Nuclear Research. We both thank Ross Young and Jose Goity for
interesting discussions. {}Finally, this work was supported in part
by DOE contract DE-AC05-84ER40150, under which SURA operates
Jefferson Laboratory.\\[20pt]

\end{document}